\documentclass[12pt]{iopart}

\usepackage{epsf}
\def\degree{\char'27\kern-.3em \hbox{C} }
\def\Cd2Re2O7{Cd$_{2}$Re$_{2}$O$_{7}$}
\def\A2B2O6O'{A$_{2}$B$_{2}$O$_{6}$O'}
\def\Tc{$T_{\rm c}$}
\def\Hc2{$H_{\rm c2}$}

\begin{document}

\title{Superconductivity in a pyrochlore oxide \Cd2Re2O7}

\author{Hironori Sakai$^{1,2}$, Kazuyoshi Yoshimura$^{1}$, Hiroyuki Ohno$^{1}$,\\
Harukazu Kato$^{2}$, Shinsaku Kambe$^{2}$, Russell E. Walstedt$^{2}$,\\
Tatsuma D. Matsuda$^{2}$, Yoshinori Haga$^{2}$,\\
and Yoshichika \={O}nuki$^{2,3}$}

\address{$^{1}$ Department of Chemistry, Graduate School of Science, Kyoto University,
 Kyoto 606-8502, JAPAN}

\address{$^{2}$ Advanced Science Research Center, Japan Atomic Energy Research Institute,
Tokai, Ibaraki 319-1195, JAPAN}

\address{$^{3}$ Graduate School of Science, Osaka University, Toyonaka,
Osaka 560-0043, JAPAN}

\begin{abstract}
We make the first report  that a metallic pyrochlore oxide \Cd2Re2O7, exhibits type II 
superconductivity at 1.1 K. 
The pyrochlore oxide is known to be a geometrical frustrated system,
which includes the tetrahedral network of magnetic ions.
A large number of compounds are classified in the family of pyrochlore oxides,
and these compounds exhibit a wide variety of physical properties ranging from insulator through semiconductor and from bad metal to good metal.
Until now, however, no superconductivity has been reported for frustrated pyrochlore oxides.
The bulk superconductivity of this compound is confirmed
by measurements of the resistivity and the a. c. magnetic susceptibility.
The \Hc2, which is extrapolated to 0 K, is estimated as about  0.8 T,
using the resistivity measurements under aplied field.
The plot of \Hc2 vs $T$ indicates that the Cooper pairs are composed of rather heavy quasiparticles.
This fact suggests that frustrated heavy electrons become superconducting in this compound.
\end{abstract}



\maketitle

Recently the subject of geometrical frustration in strongly correlated electron systems has attracted considerable interest.
The ground states of these systems are expected to be highly degenerate.
Such high degeneracies lead to thermodynamic instability at low temperatures.
Lifting these degeneracies makes possible the production of exotic quantum ground states, such as spin-liquid, heavy fermion, and unconventional superconductivity.

\begin{figure}
\begin{center}
\epsfbox{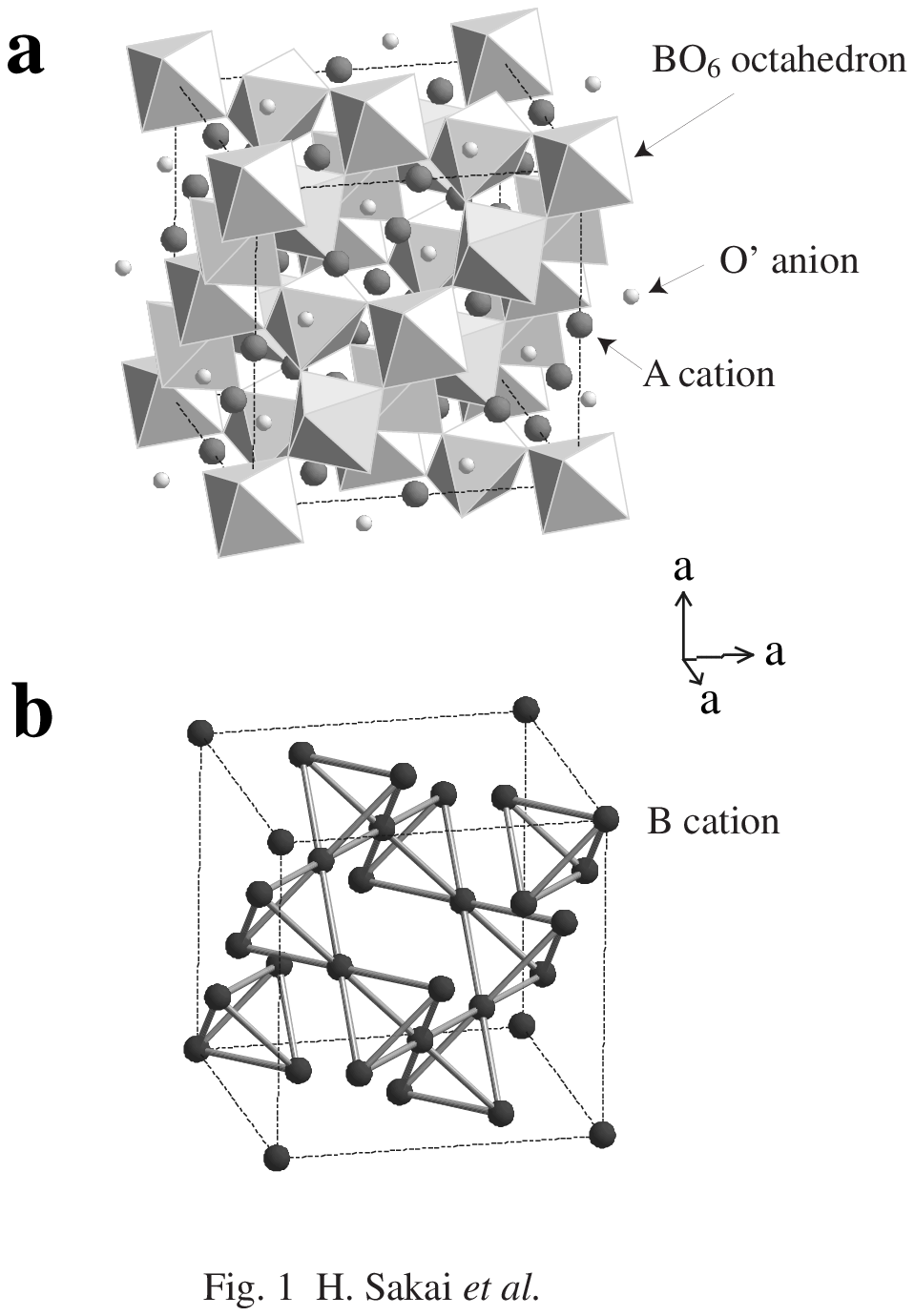}
\end{center}
\caption{\label{CrystalStructure}Crystal structure of pyrochlore oxide. 
Figure 1a is drawn on a basis of 
the network of BO$_{6}$ octahedra.
Figure 1b shows the pyrochlore lattice of B cations.}
\end{figure}

The pyrochlore oxide, which has a general formula \A2B2O6O', contains a tetrahedral network of  A or B cations, leading to geometrical frustration.
The A cations are eight-fold coordinated with six O and two O' anions and are located within distorted cubes, while the smaller B cations are six-fold coordinated and are located within distorted octahedra of which the six bond lengths from the central B cation to the corner O anion are equal.
The corner-sharing BO$_{6}$ octahedra compose a three-dimensional network as shown in Fig. \ref{CrystalStructure}a.
The sublattice of B cations composes a three-dimensional corner-shared tetrahedral network, namely, the pyrochlore lattice as shown in Fig.\ref{CrystalStructure}b.
If the A cations are non-magnetic and the B cations are magnetic, the B-spin magnetic couplings are strongly frustrated under a nearest-neighbor antiferromagnetic exchange interaction.
For the case of localized electron spin moments, theoretical studies of the Heisenberg model on the pyrochlore lattice have suggested that the ground state of such an insulator would be long-range magnetic order\cite{Anderson}, spin-freezing\cite{Villain}, or quantum spin liquid\cite{Reimers}.
The geometrical frustration plays a crucial role even in the case of itinerant electrons in the pyrochlore lattice. 
Indeed, the metallic spinel oxide LiV$_{2}$O$_{4}$, which has a pyrochlore lattice of vanadium, has been reported to show heavy fermion behavior at low temperatures due to the geometrical frustration\cite{SKondo}.

A large number of compounds are classified in the family of pyrochlore oxides, and these compounds exhibit a wide variety of physical properties ranging from insulator through semiconductor and from bad metal to good metal\cite{review}.
Until now, however, no superconductivity has been reported for frustrated pyrochlore oxides.
Here, we make the first report that a metallic cubic pyrochlore oxide \Cd2Re2O7, exhibits type II superconductivity at 1.1 K. 

Polycrystalline samples of \Cd2Re2O7 were prepared by solid state reaction. A 
stoichiometric mixture of CdO, ReO$_{3}$, and Re metal was pelletized, and put into an 
alumina Tammann tube. The pellet in the Tammann tube was further inserted into an 
evacuated silica tube and preheated at 300 \degree for several hours in order to avoid a 
vaporization of the starting materials. Then, the pellet was heated at 1000 \degree for several 
hours. The powder XRD pattern measured at room temperature was identified as the 
cubic pyrochlore structure with a lattice parameter a=10.221\AA, which is consistent 
with the previous report\cite{Donohue}.

\begin{figure}
\begin{center}
\epsfbox{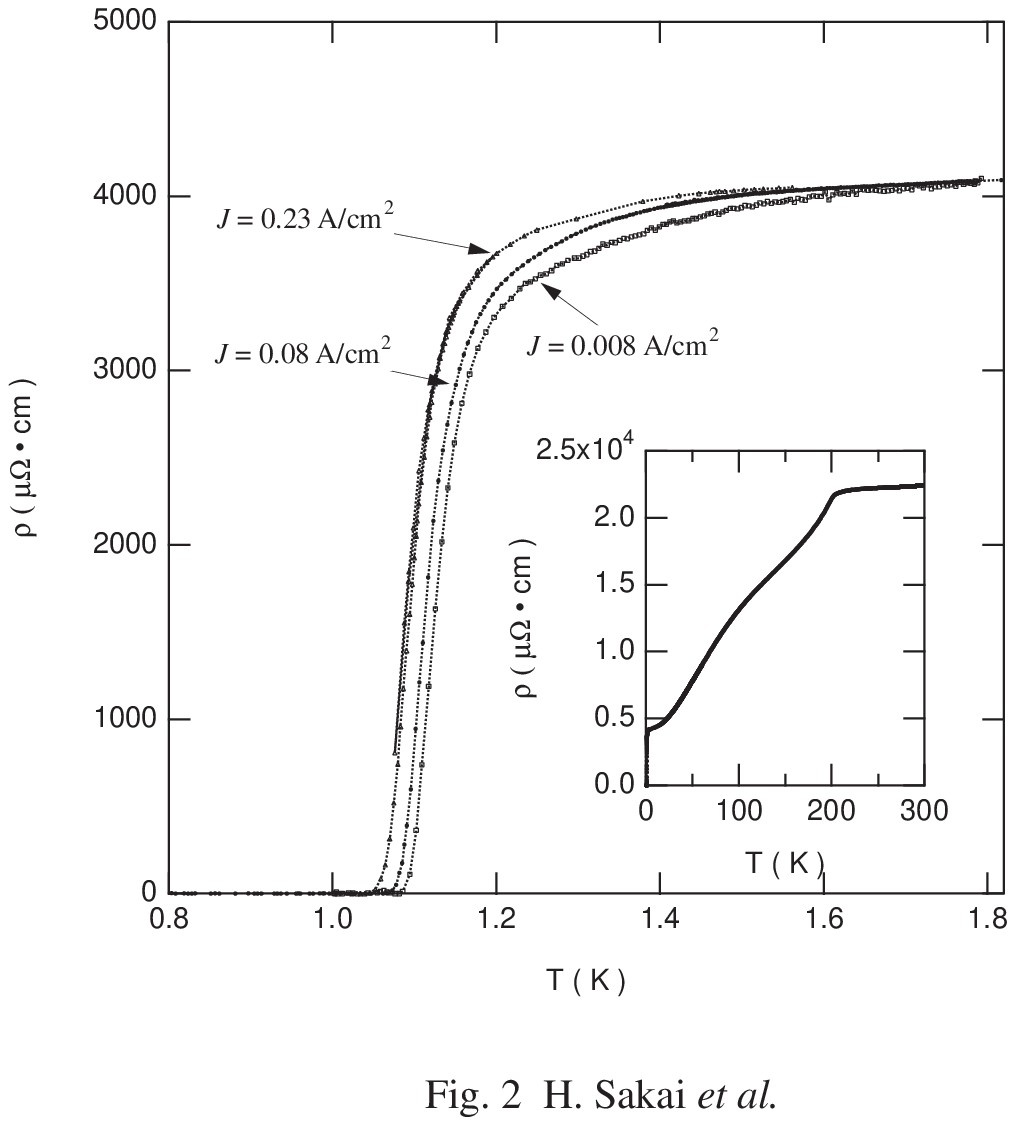}
\end{center}
\caption{\label{resistivity}Temperature dependence of d. c. electrical resistivity
for \Cd2Re2O7 in the temperature range of 0.8 to 1.8 K.
The inset shows the data in the range of 0.3 to 300 K.}
\end{figure}

The direct current (d. c.) electrical resistivity of the sintered sample of \Cd2Re2O7 
was measured using a standard four-probe technique in the temperature range of 0.3 to 
300 K under applied field from 0 to 2 T.
Low-temperature measurements below 1.9 K were performed using a $^{3}$He refrigerator.
Figure \ref{resistivity} shows the temperature dependence of 
the d.c. electrical resistivity of \Cd2Re2O7 under zero applied field. The electrical 
resistivity shows a steep descent below $T^{*}\sim$ 200 K. This anomaly at $T^{*}$ was observed in the d.c. magnetic susceptibility measurement as well. 
The origin of this anormaly has not been identified.
The electrical resistivity drops to zero sharply at the onset superconducting temperature \Tc =1.1 K, and shows effectively zero resistivity below 1.05 K. Changes of driving electric current density produced slight differences in the resistivity below 1.7 K, an effect which may be due to the superconductivity of a small amount of impurity rhenium metal (\Tc =1.7 K). 
The superconductivity transition at 1.1 K is not due to filamentary superconductivity of Re, 
since the observed critical field is much larger than the critical field of Re (0.02 T), as 
will be described below. The large residual resistivity of $4\times 10^{-3} \Omega\cdot$cm may suggest a low carrier density of this system. From band calculations, it is pointed out that the 
system has a very small density of states at the Fermi level, which is located within the 
5d-band and in the valley between the very flat bands of the rhenium 5d electrons\cite{Harima}. 
However, the electronic specific heat coefficient has been found to be large, {\it i. e.}, $\gamma$=13.3 
mJ/Re mol$\cdot$ K$^{2}$\cite{Blacklock}, indicating that heavy quasiparticles are formed due to the spin-frustration in this compound.

\begin{figure}
\begin{center}
\epsfbox{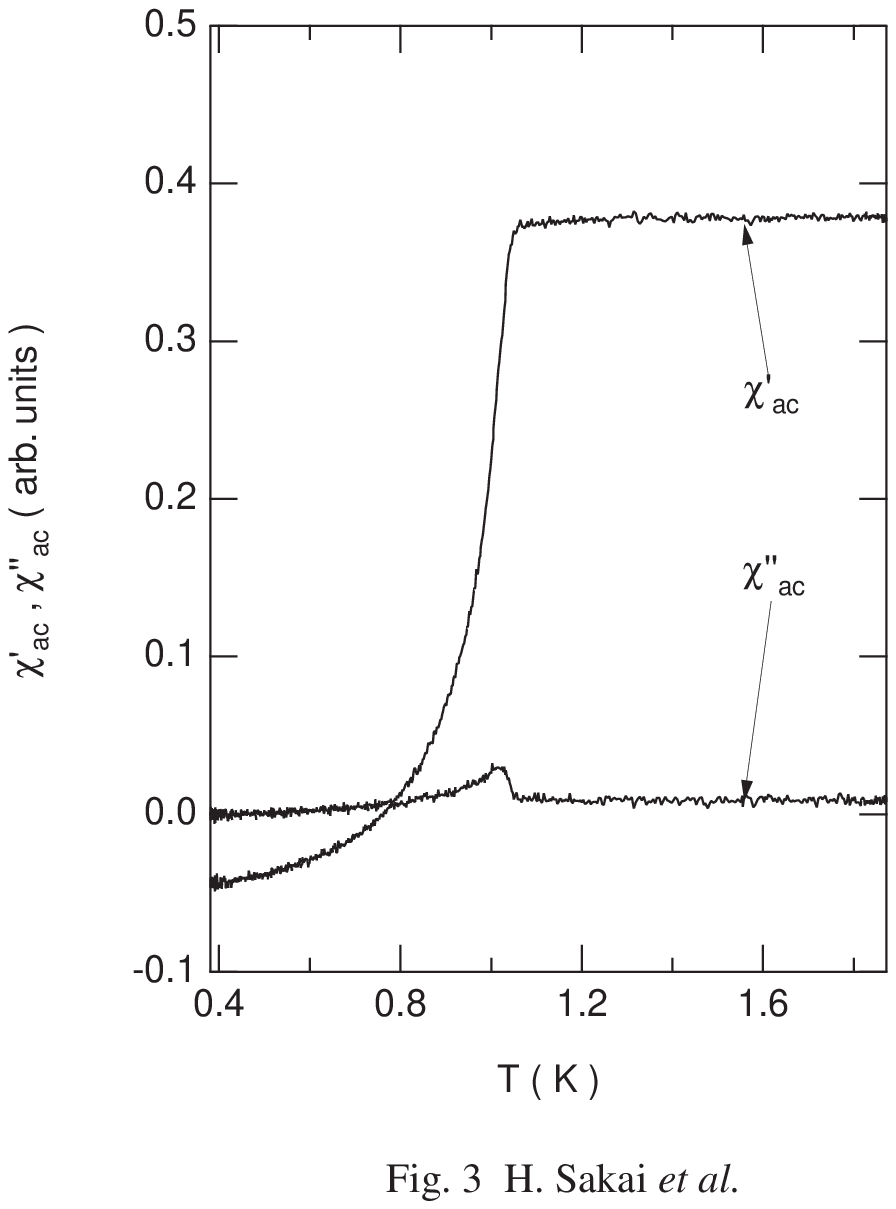}
\end{center}
\caption{\label{acChi}Temperature dependence of a.c. magnetic susceptibility for \Cd2Re2O7. 
$\chi$' and $\chi$'' represent the real part and the imaginary part, respectively.}
\end{figure}

The alternating current (a. c.) magnetic susceptibility was measured by a mutual-
inductance method at a magnetic field of 2$\times$10$^{-5}$ T
in the temperature range of 0.3 to 1.8 K.  
As shown in Fig. \ref{acChi}, a strong diamagnetic signal ($\chi$') has been observed below 1.06 K, 
which corresponds to the end-point transition temperature of the electrical resistivity 
measurement. The dissipative component $\chi$'' shows only a small peak around \Tc, 
indicating no weak superconducting link between superconducting grains. The 
superconducting volume fraction was estimated roughly as $\sim$ 50 \% at 0.3 K. From this 
experiment, a bulk superconducting state has been strongly confirmed in \Cd2Re2O7 
below 1.1 K.

\begin{figure}
\begin{center}
\epsfbox{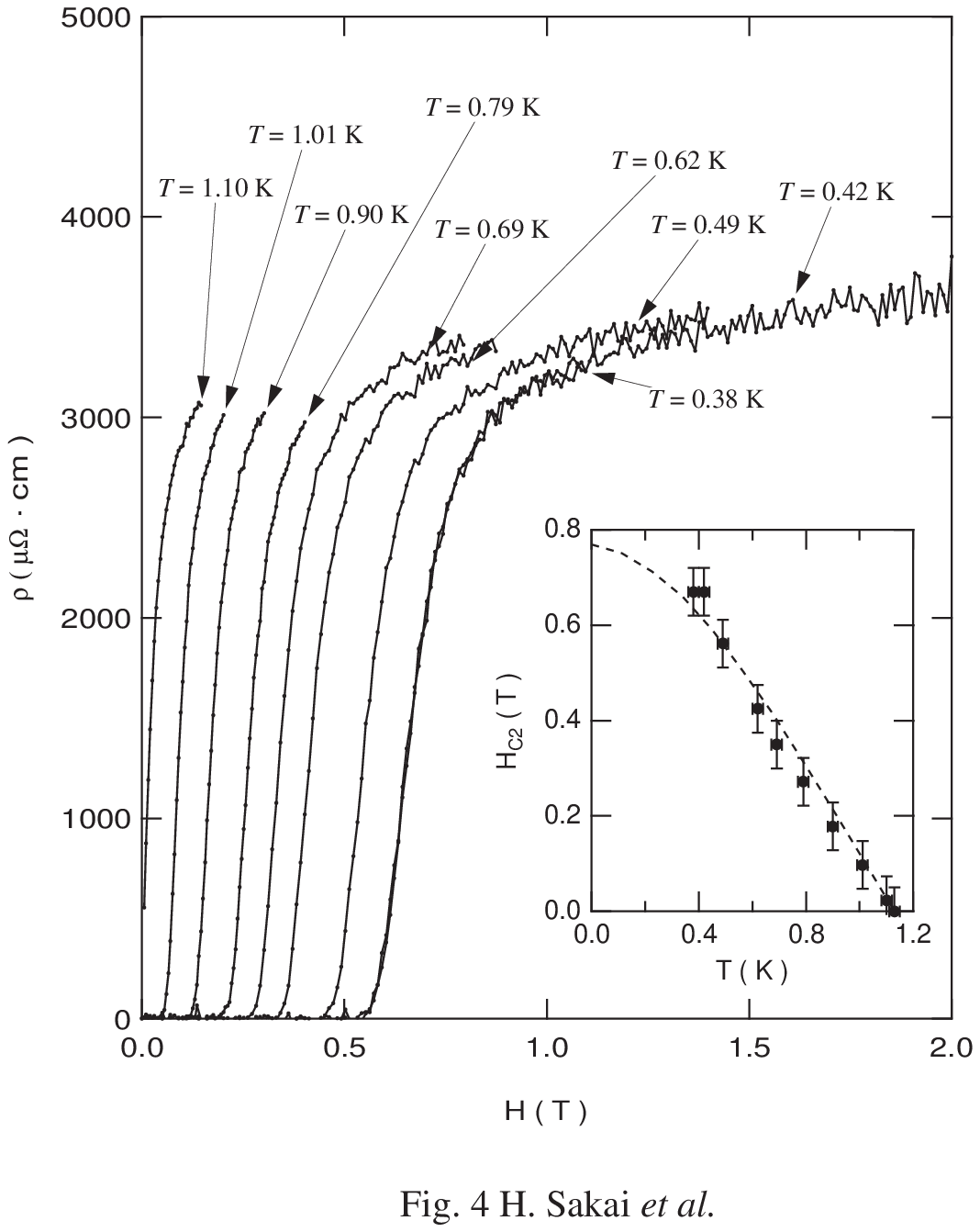}
\end{center}
\caption{\label{Hc2}Magnetic field dependence of the resistivity for \Cd2Re2O7.
The inset shows the plot of \Hc2 vs $T$.}
\end{figure}

To estimate the superconducting critical field, the magnetic field dependence of 
\Tc has been determined by resistivity measurements.
As the magnetic field was applied, the sample maintained zero resistivity until the field reached \Hc2.
At \Hc2 the superconductivity was quenched abruptly as shown in Fig.\ref{Hc2}.
The value of the upper critical field, which is extrapolated to zero temperature in the plot of \Hc2 vs $T$ (Fig. \ref{Hc2}, inset) using the WHH model\cite{WHH}, is estimated as \Hc2 (0)=0.8 T.
The superconducting coherence length,  $\xi$ is expressed as  $\sqrt{\frac{\phi _{0}}{2\pi H_{\rm c2}(T)}}$ (where $\phi_{0}$ is the fluxoid quantum). 
The Pippard coherence length, $\xi _{0}=\frac{\hbar v_{\rm F}}{\pi\Delta(0)}$ (where $v_{\rm F}$ is the Fermi velocity, and  $\Delta(0)$ is the superconducting energy gap at $T$=0 K) was obtained as about 300 \AA\ using the relation, $\xi(T)=\frac{0.74\xi_{0}}{\sqrt{1-\frac{T}{T_{\rm c}}}}$  for a clean superconductor.
This compound is considered to be a clean superconductor,
since the mean free path $l$ is estimated as
$\sim 4\times 10^{-8}$ m $>$ $\xi_{0}$, using the Drude relation,
$\rho=\frac{\hbar(3\pi^{2})^{\frac{1}{3}}}{e^{2}l}n^{-\frac{2}{3}}$
with resistivity $\rho\sim 4\times 10^{-5}\ \Omega\cdot$ m
and carrier concentration $n\sim 1\times 10^{24}$ m$^{-3}$ which is a typical value for metallic pyrochlore compounds.
The observed large initial slope of 
\Hc2, {\it i.e.}, $\Bigl(\frac{1}{T_{\rm c}}\frac{dH_{\rm c2}}{dT}\Bigr)_{T=T_{\rm c}}$ $\sim$ 1 T/K$^{2}$, indicates that the Cooper pairs are composed of rather 
heavy quasiparticles, since $\Bigl(\frac{1}{T_{\rm c}}\frac{dH_{\rm c2}}{dT}\Bigr)_{T=T_{\rm c}}$ is proportional to the square of effective 
mass $m^{*2}$, in agreement with the large $\gamma$ value.
This fact suggests, indeed, that frustrated 
heavy electrons become superconducting in this compound.

Generally, superconductivity due to a magnetic interaction competes with 
magnetic frustration, since no particular magnetic excitation for a superconducting 
attractive channel exists due to the frustration. The present study gives the first example 
of coexistence of magnetic frustration and superconductivity.

\end{document}